\documentclass[%
 aps,
 prl,%
 amsmath,amssymb,
 reprint,%
 superscriptaddress,
]{revtex4-2}
\usepackage{xcolor}
\usepackage{graphicx}
\usepackage{dcolumn}
\usepackage{bm}
\usepackage[mathlines]{lineno}

\begin{document}

\preprint{AIP/123-QED}

\title{Shaping the angular spectrum of a Bessel beam to enhance light transfer through dynamic strongly-scattering media}
\author{Dennis Scheidt}%
\email{dennis.scheidt@correo.nucleares.unam.mx}
 \affiliation{ 
Instituto de Ciencias Nucleares, Universidad Nacional Aut\'onoma de M\'exico, Apartado Postal 70-543, 04510, Cd. de México, México
}
\author{Alejandro V. Arzola}%
\affiliation{Instituto de Física, Universidad Nacional Autónoma de
México, C.P. 04510, Cd. de México, México}

\author{Pedro A. Quinto-Su}%
 \email{pedro.quinto@nucleares.unam.mx}
\affiliation{ 
Instituto de Ciencias Nucleares, Universidad Nacional Aut\'onoma de M\'exico, Apartado Postal 70-543, 04510, Cd. de México, México
}

\begin{abstract}
\noindent We prepare a quasi-non-diffracting Bessel beam defined within an annular angular spectrum 
with a spatial light modulator. The beam propagates though a strongly scattering media and the transmitted speckle pattern is measured at one point with a  Hadamard Walsh basis that divides the ring into $N$ segments ($N=16,64,256, 1024$). The phase of the transmitted beam is reconstructed with 3 step interferometry and the intensity of the transmitted beam is optimized by projecting the conjugate phase at the SLM. We find that the optimum intensity is attained for the condition that the transverse wave vector $k_\perp$ (of the Bessel beam) matches the spatial azimuthal frequencies of the segmented ring $k_\phi$. Furthermore, compared with beams defined on a 2d grid (i.e. Gaussian)  a reasonable enhancement is achieved for all the $k_\perp$ sampled with only 64 elements. 
Finally, the measurements can be done while the scatterer is moving as long as the total displacement during the measurement is smaller than the speckle correlation distance.
\end{abstract}


\maketitle

\noindent Control and manipulation of light through diffusive media has many applications in physics and biology, for instance, to tightly focus light inside tissue to trap cells in-vivo \cite{insitu1,OpticalTweezersCorrection, vcivzmar2011shaping}, to image through diffusive media \cite{ohta20223d,han2023extending, harm2014lensless,dalgarno2012wavefront,andrews2005laser} or to tailor structured beams \cite{yuan2021experimental,di2016tailoring,wang2016deep,chen2018needle,dalgarno2012wavefront}.  
To prevent the diffusion of light by a diffusive media or to recover the undisturbed wave front, it is necessary to recover the complex transmission function of the media \cite{T_Mat1, T_Mat3, Skarsoulis:21, insitu1}. This is commonly accomplished with an Spatial Light Modulator (SLM) following wave front shaping protocols.

The transmitted light can be optimized at one point by means of wavefront correction techniques that resolve the phase of each mode of the entering beam to enhance the intensity \cite{insitu1}. In this approach, the modes of the beam are defined by the elements of a 2D grid of the transverse profile of the beam in the 
frequency space.  The phase of each element of this 2D basis (canonical basis) can be independently found in order match the phases of all modes, optimizing in this way the overall intensity of the beam after transmission through the scattering medium \cite{insitu1}. Recently, it has been shown that Hadamard-Walsh basis in contrast to the canonical basis drastically improves the correction of the diffused beams, making it feasible to optimize the transmission of light in conditions where the canonical basis commonly fails \cite{OurPaper}.


Bessel beams are a type of invariant beams widely used in a broad range of applications -- including optical trapping, imaging and laser cutting \cite{mcgloin2005bessel,fahrbach2013light,Fahrbach:12}. In practice, a close description to these beams are the Gauss-Bessel beams, having finite energy and finite-length propagation \cite{mcgloin2005bessel, forbes2014laser}. These beams have unique properties such as the ability to propagate very long distances almost without diffraction and to avoid obstacles by means of the self-healing phenomenon \cite{SelfHealing_review, Fahrbach:12}. These also allow to generate narrower spots than Gauss beams \cite{mcgloin2005bessel}. 
Another important property of these beams is the shape of the power spectrum, which is defined by a ring of radius equal to the transverse component of the wave vector in frequency space \cite{forbes2014laser}.


Here, we use the angular spectrum of a Bessel beam to control the propagation of light trough strongly scattering media. Following Durnin's approach to generate a Bessel beam \cite{DurninRingBessel, durnin1987diffraction}, a ring of light sculpted by a SLM propagates trough a lens to generate an approximately-invariant Bessel beam near the focal region, which is strongly diffused by a ground-glass diffuser located near the focal position. To enhance transmission of light at one point, we segment the ring-shaped spectrum located in the SLM to build a 1D basis. We sample the transmitted phase by means of the Hadamard Walsh basis and three-step interferometry. We also explore the capability of this protocol to control transmission of light through a dynamic diffuser. For this purpose, firstly, we characterize the degradation of the optimized transmitted beam as the diffuser is slowly moved from its original position. This allows us to determine a characteristic displacement where the optimized transmitted light is completely lost. With this in hand, finally, we show a protocol to optimize the transmission of light through a dynamical strongly-scattering medium (moving ground-glass diffuser).

\begin{figure}
    \centering
    \includegraphics[width = 0.40\textwidth]{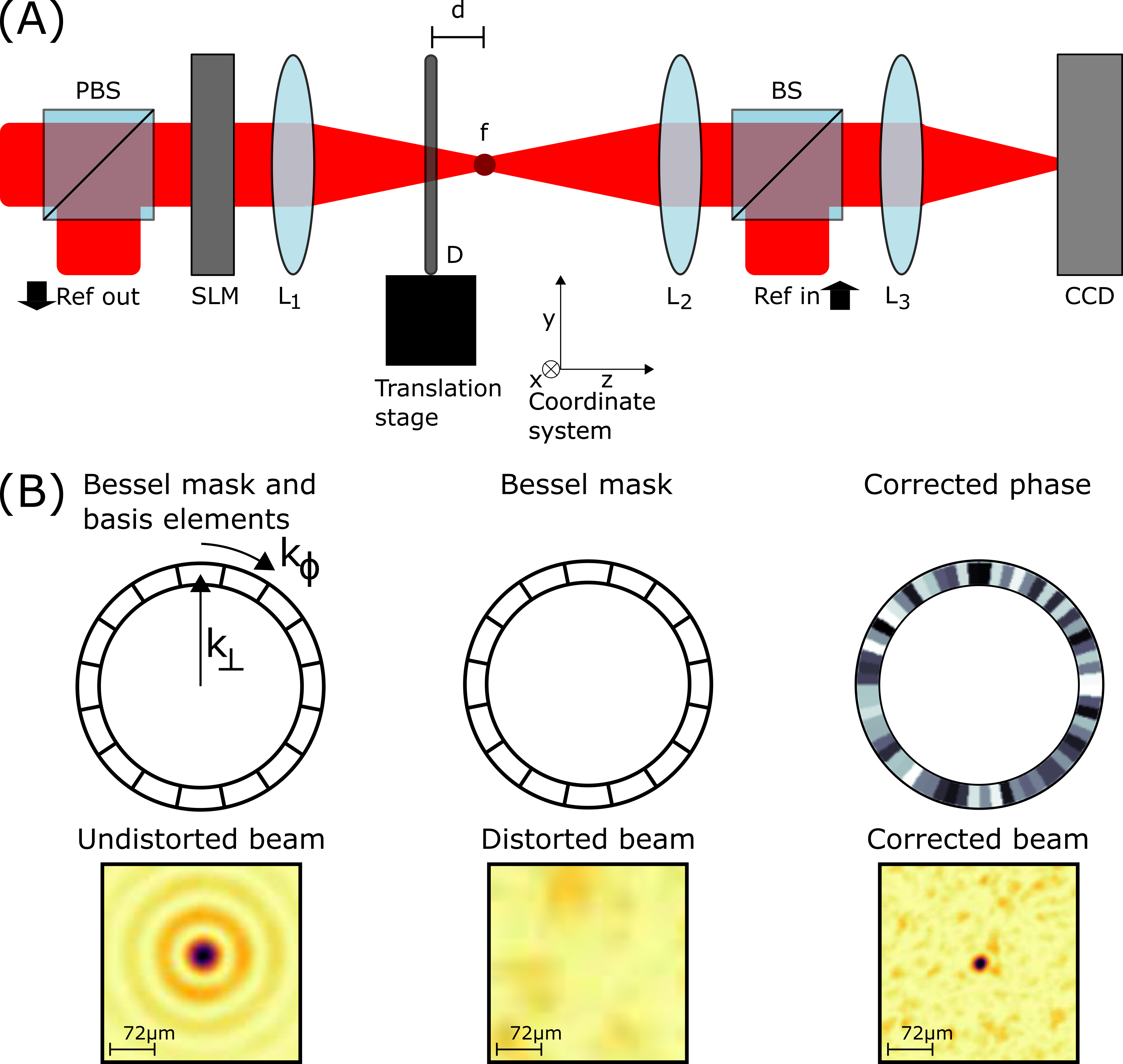}
    \caption{Experiment setup and ring basis for Bessel beam. A) Schematic of the setup. Optical elements: Polarizing beam splitter (PBS), beam splitter (BS), spatial light modulator (SLM), lenses $\text{L}_i$, diffuser (D). The position of unperturbed focal point is indicated by  f. B) Implementation of the ring basis on the SLM and assignment of the corrected phase with the resulting beam profiles. $k_\perp$ is the spatial frequency of the Bessel beam defined by the ring mask and $k_\phi$ is the spatial frequency of the basis elements. The displayed windows span over $3.6$ mm.
    }
    \label{fig:fig2}
\end{figure}

\vspace{0.3cm}
\noindent {\it Experiment.} 
The experimental setup is schematically shown in Fig.\,1. and is similar to the one described in \cite{OurPaper}. 
A linear polarized and expanded HeNe laser ($\lambda = 633\,$nm, $P = 0.395 \pm 0.02\,$mW) is split up into a reference (vertical polarization state) and modulated beam (horizontal polarization state) with a polarizing beam splitter (PBS). 
The horizontal component (modulated beam) is reflected at the surface of a SLM (SLM - Hamamatsu: LCOS-SLM X10468) with a beam waist of $1.50$cm. 
The beam is focused with $L_1$ ($f_1=500\,$ mm), recollimated with $L_2$ ($f_2=150\,$ mm) and then refocused with $L_3$ ($f_3=200\,$ mm) onto the CCD.
The beam is distorted using a glass diffuser $D_1$ (Thorlabs: DG10-120/600-MD) which is placed $d = 1\,cm$ behind the geometrical focus of $L_1$. 
The reference beam is controlled with a motorized shutter (not shown) and propagates through a half-wave plate (not shown) to match the polarization of the modulated beam before it is combined the modulated beam in a second beam splitter (BS).
A lens ($L_3$, $f_3 = 200$ mm) focuses the interfering beams into a CCD camera (Thorlabs DCC1645C, $3.6\,\mu$m pixel size). 

{\it Bessel beam.} 
The Bessel beam is created using the method described by Durnin \cite{BesselBeams} by projecting a ring of radius $r_0$ with a width of $\Delta r$ onto the surface of the SLM and using a lens L$_1$ to produce the Fourier transform of the ring illumination, resulting in a Bessel beam with a spatial frequency of:
\begin{equation}
    k_\perp = k\cdot \sin\left(\frac{r_0}{f}\right) \ ,
\end{equation}
\noindent where $k=2\pi/\lambda$ and the focal length $f$ of the focusing lens L$_1$. In our experiment we keep a fixed ring width $\Delta r = 0.72$ mm and vary $r_0$ ($k_\perp$) in the range of $[1.12,5.32]$\,mm.
As a result, different depth of foci (DOF) of diffraction free length are achieved according to \cite{DurninRingBessel}:
$DOF = \frac{2.8 f^2}{r_0 \Delta r k}$.

The beam is transmitted through a random scatterer resulting in a speckle pattern.
In order to measure the effective phase induced by the scatterer at different sections of the ring we use a Mach-Zehnder interferometer and the phase is sampled with a Hadamard Walsh basis.

{\it Ring Basis.}
The field is represented with the Hadamard-Walsh basis HW, which forms a orthogonal $N\times N$ matrix ($N = 2^m, m \in \mathbb{N}$) consisting of $+1$ and $-1$ entries. The first column of Fig. 1B has a segmented ring that is projected on the SLM, where the elements are assigned values of $\pm 1$. The photograph in at the bottom is the imaged unperturbed Bessel beam.

The Walsh ordering of the Hadamard matrix is based on increasing spatial frequency. Hence the first vectors have the lowest spatial frequency and the last the highest. In order to measure a $1\times N$ basis vector $H_i$, the ring is divided into $N$ angular segments. Separate sampling of the '$+1$' and '$-1$' entries is avoided by assigning a value of '$+\pi$' to the '$-1$' segments. 

Similar to the transverse spatial frequency  of the Bessel beam given by $k_\perp$, we can define the azimuthal spatial frequency of the ring basis $k_\phi$:
\begin{equation}
    k_\phi = \frac{2 \pi N}{2 \pi r_0} = \frac{N}{r_0}
\end{equation}
\noindent with $N$ the number of elements in the full basis. 
The resulting Bessel beam has a spot size that is inversely proportional to $k_\perp$. 
Assuming that we can not measure a smaller spatial resolution than the one given by $k_\perp$, then the condition $k_\perp \sim k_\phi$ yields. 
\begin{equation}
    N_{opt} = \frac{k}{f} r_0^2 \ ,
\end{equation}
where $N_{opt}$ represents the optimal number of basis elements so that the transverse and azimuthal frequencies match. A smaller basis won't have sufficient resolution while a larger one won't improve the measurements.
Then we have to round $N_{opt}$ to the discrete sets of sizes of the Hadamard basis $N=4^n$ (n is an integer):
\begin{equation}
    N=4^{\text{ceil} (\log _4 (N_{opt}))}
\end{equation}
Which is the basis size chosen for the Hadamard basis as a function of $k_\perp$ and might oversample when $N_{opt}$ is smaller than the rounded value in eq. (4).

\vspace{0.5cm}
{\it Measurement and optimization.}
The wavefront correction is achieved in two steps: First, we measure the phase of the transmitted field and then apply the conjugated phase as a second step to measure the signal yield.
The middle column of Fig. 1B contains the ring that makes the Bessel beam and the bottom one is the photograph of the perturbed Bessel. The last column contains the corrected phase ring and the imaged spot. 
Note that the unperturbed Bessel mask contains $N=16$ elements, while the corrected phase mask has $N =64$ basis elements.
Notice that the corrected spot is smaller than the unperturbed beam, which is an effect of disordered media \cite{Vellekoop_2010_localization}.

\vspace{0.5cm}
The measurement of the complex contribution of each basis vector $H_i$ is done with 3-step interferometry by adding a constant phase offset of $\Delta \phi_m = [0,2/3\pi,4/3\pi]$ ($m=1-3$) to the modulated beam on the SLM screen and measuring the resulting intensity $I_{i,m}$ at one point.
After iterating over the whole HW basis, each phase contribution is reconstructed with $x_m = H^{-1} \cdot I_m$. 
The complex light field is retrieved by combining the fields for every offset \cite{Zupancic:16}:
\begin{equation}
    x = -\frac{1}{3}(x_{2} + x_{3} - 2x_{1}) + \frac{i}{\sqrt{3}} (x_{2} -x_{3}) \ .
\end{equation}
Calculating the argument of the complex field $\mathrm{arg}(x)$ yields the measured phase distribution and projecting the conjugate of the phase onto the SLM cancels the aberrations. The quality of enhancement is measured using the signal to noise ratio defined as: 
\begin{align}
    \text{SNR} = \frac{I_{corr}^{max}}{\hat{I}_{uncorr}} \ ,
\end{align}
\noindent where $\hat{I}_{uncorr}$ is the average value of the uncorrected and $I_{corr}^{max}$ the maximal value of the corrected image taken with the CCD camera, respectively. 

\begin{figure}
    \centering
    \includegraphics[width = 0.40\textwidth]{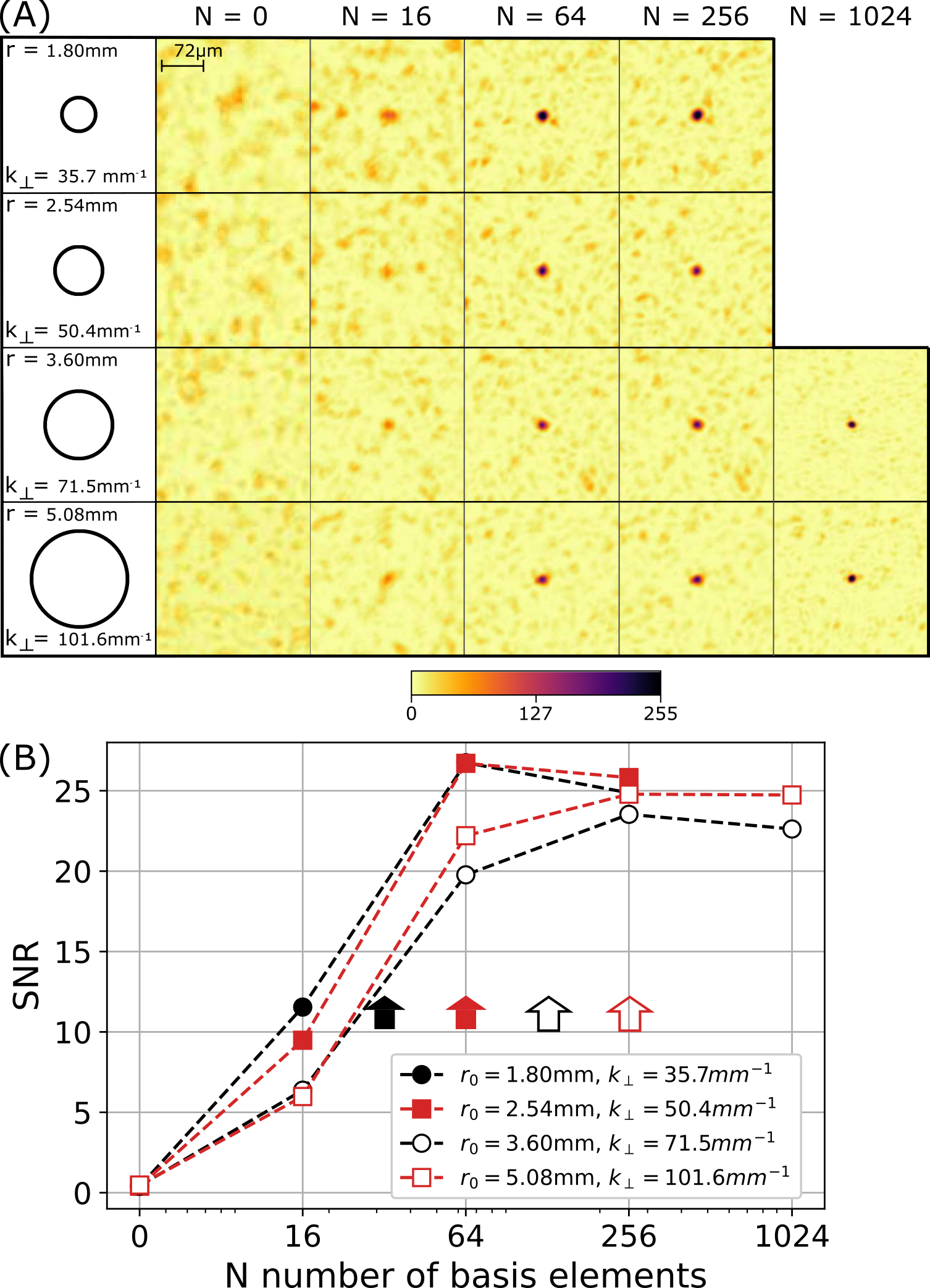}
    \caption{Bessel beam correction depending on the number of basis elements. 
    (A) Corrected Bessel beams with different $k_\perp$ depending on the number of basis elements $N$.
    (B) Signal to noise ratio of the beams.}
\end{figure}

\vspace{0.3cm}
{\it Results.} Images of the optimized spots for different $k_\perp$ as a function of basis elements ($N$, $k_\phi$) are shown in Fig. 2A. The first column has the information about the ring size ($r_0 = [1.8,\, 2.54,\, 3.6,\, 5.08]$ mm) and the corresponding $k_\perp$ ($k_\perp = [35.7,\, 50.4,\, 71.5,\, 101.6]\text{ mm}^{-1}$). The following columns contain the results for $N=0,\, 64,\, 256,\, 1024$, which correspond to  $k_\phi = N/r_0$. As a reference, the uncorrected speckle pattern is presented on the $N=0$ column. In the case of the smaller $k_\perp$ there is no $N=1024$ because the width of the ring slices is smaller than 1 pixel at the SLM.

The different behaviours of the SNR for each corrected spot as a function of $N$ are in Fig. 2B. The filled circles and squares represent $k_\perp =35.7, \, 50.4\text{ mm}^{-1}$ respectively, while the unfilled symbols correspond to $k_\perp =71.5, \, 101.6\text{ mm}^{-1}$. The optimum values for the number of basis elements for each $k_\perp$ are calculated with eq. (3) and yield $N_{opt} = [32,\,64,\,128,\,256]$. The arrows in the plot (Fig. 2B) show the position of $N_{opt}$ for their respective $k_\perp$. In the case of $k_\perp =35.7, \,71.5\text{ mm}^{-1}$ the $N_{opt}=32,\, 128$ are between the permitted values of $N=4^n$ for the HW basis and as expected, the largest SNR are attained at the rounded $N=64, \,256$ (eq. (4)) that oversamples the optimum number of elements. In the case of $k_\perp =50.4, \,101.6\text{ mm}^{-1}$ the $N_{opt}=64, \,256$ coincides with the permitted sizes of the HW bases and yield the largest SNR.

Interestingly, we observe that in all cases the largest increase in SNR is that between $N=16$ and $N=64$, so 64 elements already yield a very good enhancement.

\begin{figure}
    \centering
    \includegraphics[width = 0.40 \textwidth]{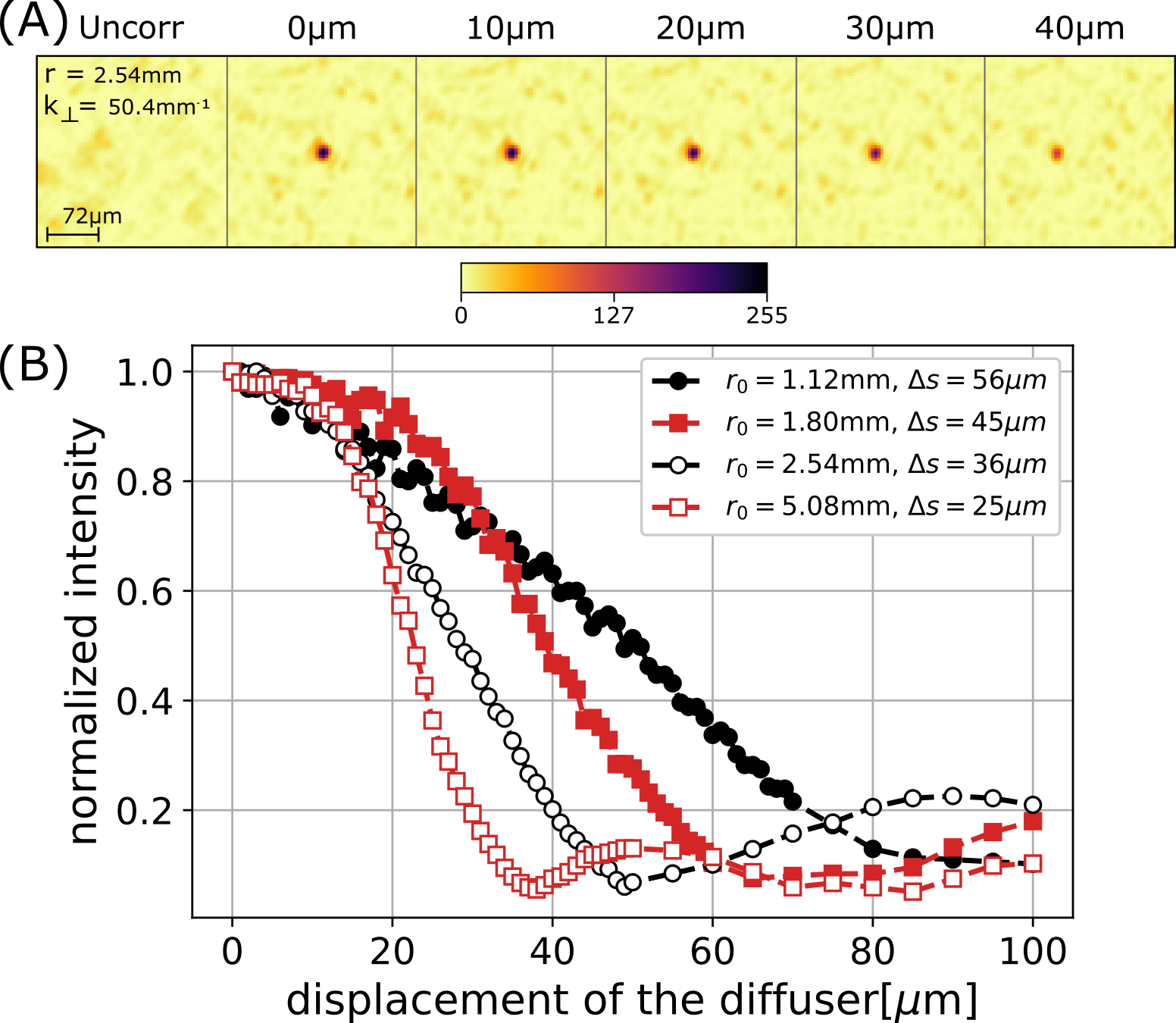}
    \caption{Displacement of the diffuser after correction.
(A) Photographs of the uncorrected and corrected spot at different displacements of the diffuser.
(B) Normalized intensity of the corrected spot after diffuser displacement $\Delta x$. $\Delta s$ is the correlation distance of each Bessel beam.}
\end{figure}

\vspace{0.3cm}
{\it Speckle correlation distance measurement.}
We measure the speckle correlation distance for each $k_\perp$ by moving the diffuser in the transverse direction with a motorized translation stage (Model Newport LTA-HS Actuator Stage) in steps of $\Delta x = 1 \,\mu$m. This is done after the correction is applied with the optimum $N$. For each motion step, an image of the corrected spot is taken as shown in Fig. 3A 
for $k_\perp=50.4\text{ mm}^{-1}$ ($N=64$). 
Notice that the spot moves along the translation direction (horizontal direction).
Therefore the normalized maximum intensity of the spot is plotted as a function of the  diffuser displacement in Fig. 3B for  $k_\perp = [50.4, \, 71.5, \, 101.6]\text{ mm}^{-1}$.
We define the correlation distance $\Delta s$ as the distance where the intensity drops to $1/e$ \cite{goodman2020speckle}. We observe a reciprocal relationship between $k_\perp$ and the correlation distance $\Delta s$. This correlates with the Bessel beams spot size of the main lobe which is also inverse proportional to $k_\perp$.
The results suggest that as long as we do not exceed that distance, the diffuser can move during the measurement.

\vspace{0.3cm}
{\it Moving ground-glass diffuser.} In order to put the method to test in dynamically changing environments, we displace the diffuser with different constant velocities ($0.1, 0.52, 1.04, 1.56\,\mu$m/s) along the x-direction. 
We use the case of a beam with ring radius of $r_0 = 2.54$ mm ($k_\perp = 50.4 \text{ mm}^{-1}$), with $N=N_{opt} = 64$ and $\Delta s = 35\,\mu$m for this experiment. 

As the diffuser is moving the interferometric measurement is performed, and then we project the corrected spot (blocking the reference beam with a motorized beam block). 
This process takes $t_{acq} \approx 45$ seconds. 
As a result, the total displacement of the diffuser during a single correction results in $\Delta x = v\cdot t_{acq}$. 
The images in Fig. 4A show several images of the corrected spot at different repeated experiments while the diffuser is moving.
The first row is for $v=0.1\,\mu$m/s and shows the results in steps of 10 experiments.
The spots vary only sightly in the intensity and shape, due to the randomness of the diffuser surface.

The last row is for the largest speed ($v=1.56\,\mu$m/s) where it is not possible to make a dynamic correction because the displacement of the diffuser exceeds the speckle correlation distance.

The SNR as a function of relative displacement of the diffuser ($\Delta x_0 = \Delta x/\Delta s$) is in Fig. 4B.
The corrected intensity decreases slowly for small displacements. Once the displacement is larger than half the correlation distance, the intensity is decreased by a factor of two. 
When the full correlation distance is covered, the corrected spot is not distinguishable from the speckle intensities.

\begin{figure}
    \centering
    \includegraphics[width = 0.40\textwidth]{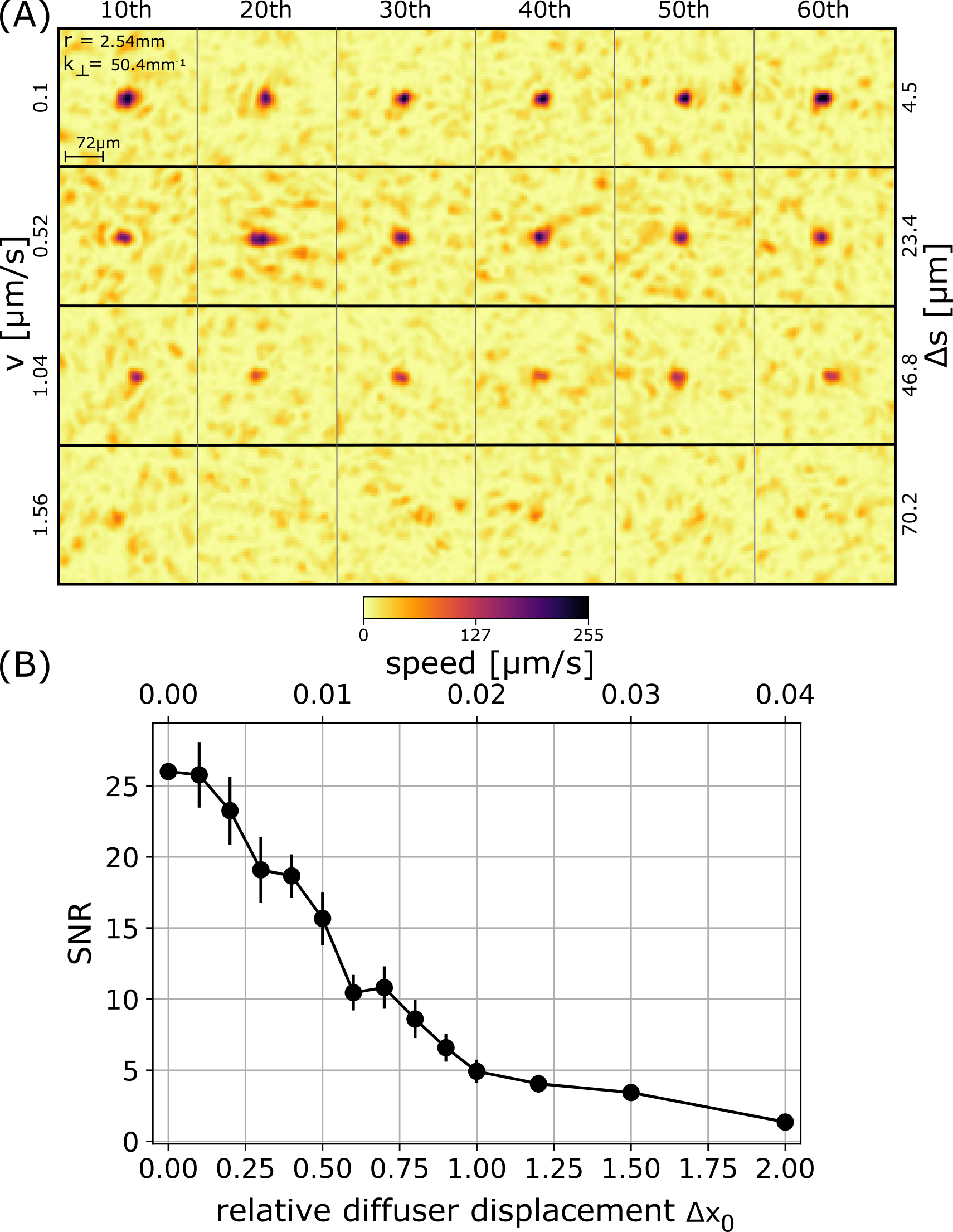}
    \caption{
    Dynamic correction through a moving diffuser.
    (A) Images of the corrected spot. The displacement speed increases from the top to bottom row from $v = 0.1\,\mu$m/s to $1.56\,\mu$m/s. With an acquisition time of $t_{acq} \approx 45$s, the total displacement distance is $\Delta x = v \cdot t_{acq}$.
    (B) SNR as a function of the relative displacement of the diffuser, being the absolute displacement $\Delta x_0 = \Delta x/\Delta s$ normalized to the correlation distance of the beam $\Delta s$. 
    }
\end{figure}

\vspace{0.3cm}
{\it Conclusion.} We found that the optimal beam correction is achieved when $k_\perp =k_\phi$. This condition allows to choose and match $k_\perp$ with the optimal segmentation of the HW ring basis ($N_{opt}=4^n$). Also, the ring basis has the advantage that it is fairly insensitive to the number of basis elements compared with that used for a regular Gaussian beam defined on a 2D grid, which increases with $N$ \cite{Vellekoop:07,OurPaper,T_Mat1}. For the range of $k_\perp$  that we explored, the optimum number of basis elements is less than or equal to $256$. In fact, $N=64$ resulted in reasonable correction for all the $k_\perp$ in our experiments. Finally, we showed that the measurements to optimize the transmission can be performed with a moving random element as long as the total displacement during the measurement does not exceed the speckle correlation distance. 

\vspace{-0.5cm}

\subsection{Funding}
\noindent Work partially funded by DGAPA UNAM PAPIIT grants IN107719 and  IN107222; CIC-LANMAC and CONACYT LN-299057. 
\vspace{-0.5cm}
\subsection{Acknowledgments} 
\noindent Thanks to Jos\'e Rangel Guti\'errez for fabricating some of the optomechanical components.
\vspace{-0.5cm}
\subsection{Disclosures} 
\noindent The authors declare no conflicts of interest.
\vspace{-0.5cm}
\subsection{Data availability} 
\noindent Data underlying the results presented in this paper may be obtained from the corresponding author upon reasonable request.

\bibliography{main_arxiv}

\begin{thebibliography}{28}%
\makeatletter
\providecommand \@ifxundefined [1]{%
 \@ifx{#1\undefined}
}%
\providecommand \@ifnum [1]{%
 \ifnum #1\expandafter \@firstoftwo
 \else \expandafter \@secondoftwo
 \fi
}%
\providecommand \@ifx [1]{%
 \ifx #1\expandafter \@firstoftwo
 \else \expandafter \@secondoftwo
 \fi
}%
\providecommand \natexlab [1]{#1}%
\providecommand \enquote  [1]{``#1''}%
\providecommand \bibnamefont  [1]{#1}%
\providecommand \bibfnamefont [1]{#1}%
\providecommand \citenamefont [1]{#1}%
\providecommand \href@noop [0]{\@secondoftwo}%
\providecommand \href [0]{\begingroup \@sanitize@url \@href}%
\providecommand \@href[1]{\@@startlink{#1}\@@href}%
\providecommand \@@href[1]{\endgroup#1\@@endlink}%
\providecommand \@sanitize@url [0]{\catcode `\\12\catcode `\$12\catcode
  `\&12\catcode `\#12\catcode `\^12\catcode `\_12\catcode `\%12\relax}%
\providecommand \@@startlink[1]{}%
\providecommand \@@endlink[0]{}%
\providecommand \url  [0]{\begingroup\@sanitize@url \@url }%
\providecommand \@url [1]{\endgroup\@href {#1}{\urlprefix }}%
\providecommand \urlprefix  [0]{URL }%
\providecommand \Eprint [0]{\href }%
\providecommand \doibase [0]{https://doi.org/}%
\providecommand \selectlanguage [0]{\@gobble}%
\providecommand \bibinfo  [0]{\@secondoftwo}%
\providecommand \bibfield  [0]{\@secondoftwo}%
\providecommand \translation [1]{[#1]}%
\providecommand \BibitemOpen [0]{}%
\providecommand \bibitemStop [0]{}%
\providecommand \bibitemNoStop [0]{.\EOS\space}%
\providecommand \EOS [0]{\spacefactor3000\relax}%
\providecommand \BibitemShut  [1]{\csname bibitem#1\endcsname}%
\let\auto@bib@innerbib\@empty
\bibitem [{\citenamefont {Cizmar}\ \emph {et~al.}(2010)\citenamefont {Cizmar},
  \citenamefont {Mazilu},\ and\ \citenamefont {Dholakia}}]{insitu1}%
  \BibitemOpen
  \bibfield  {author} {\bibinfo {author} {\bibfnamefont {T.}~\bibnamefont
  {Cizmar}}, \bibinfo {author} {\bibfnamefont {M.}~\bibnamefont {Mazilu}},\
  and\ \bibinfo {author} {\bibfnamefont {K.}~\bibnamefont {Dholakia}},\
  }\bibfield  {title} {\bibinfo {title} {In situ wavefront correction and its
  application to micromanipulation},\ }\href@noop {} {\bibfield  {journal}
  {\bibinfo  {journal} {Nature Photonics}\ }\textbf {\bibinfo {volume} {4}},\
  \bibinfo {pages} {388} (\bibinfo {year} {2010})}\BibitemShut {NoStop}%
\bibitem [{\citenamefont {Liang}\ \emph {et~al.}(2018)\citenamefont {Liang},
  \citenamefont {Cai}, \citenamefont {Wang}, \citenamefont {Lei}, \citenamefont
  {Cao}, \citenamefont {Wang}, \citenamefont {Li}, \citenamefont {Yan},
  \citenamefont {Bianco},\ and\ \citenamefont
  {Yao}}]{OpticalTweezersCorrection}%
  \BibitemOpen
  \bibfield  {author} {\bibinfo {author} {\bibfnamefont {Y.}~\bibnamefont
  {Liang}}, \bibinfo {author} {\bibfnamefont {Y.}~\bibnamefont {Cai}}, \bibinfo
  {author} {\bibfnamefont {Z.}~\bibnamefont {Wang}}, \bibinfo {author}
  {\bibfnamefont {M.}~\bibnamefont {Lei}}, \bibinfo {author} {\bibfnamefont
  {Z.}~\bibnamefont {Cao}}, \bibinfo {author} {\bibfnamefont {Y.}~\bibnamefont
  {Wang}}, \bibinfo {author} {\bibfnamefont {M.}~\bibnamefont {Li}}, \bibinfo
  {author} {\bibfnamefont {S.}~\bibnamefont {Yan}}, \bibinfo {author}
  {\bibfnamefont {P.~R.}\ \bibnamefont {Bianco}},\ and\ \bibinfo {author}
  {\bibfnamefont {B.}~\bibnamefont {Yao}},\ }\bibfield  {title} {\bibinfo
  {title} {Aberration correction in holographic optical tweezers using a
  high-order optical vortex},\ }\href {https://doi.org/10.1364/AO.57.003618}
  {\bibfield  {journal} {\bibinfo  {journal} {Appl. Opt.}\ }\textbf {\bibinfo
  {volume} {57}},\ \bibinfo {pages} {3618} (\bibinfo {year}
  {2018})}\BibitemShut {NoStop}%
\bibitem [{\citenamefont {{\v{C}}i{\v{z}}m{\'a}r}\ and\ \citenamefont
  {Dholakia}(2011)}]{vcivzmar2011shaping}%
  \BibitemOpen
  \bibfield  {author} {\bibinfo {author} {\bibfnamefont {T.}~\bibnamefont
  {{\v{C}}i{\v{z}}m{\'a}r}}\ and\ \bibinfo {author} {\bibfnamefont
  {K.}~\bibnamefont {Dholakia}},\ }\bibfield  {title} {\bibinfo {title}
  {Shaping the light transmission through a multimode optical fibre: complex
  transformation analysis and applications in biophotonics},\ }\href@noop {}
  {\bibfield  {journal} {\bibinfo  {journal} {Optics express}\ }\textbf
  {\bibinfo {volume} {19}},\ \bibinfo {pages} {18871} (\bibinfo {year}
  {2011})}\BibitemShut {NoStop}%
\bibitem [{\citenamefont {Ohta}\ \emph {et~al.}(2022)\citenamefont {Ohta},
  \citenamefont {Kodama}, \citenamefont {Miyamoto}, \citenamefont {Osten},
  \citenamefont {Takeda},\ and\ \citenamefont {Watanabe}}]{ohta20223d}%
  \BibitemOpen
  \bibfield  {author} {\bibinfo {author} {\bibfnamefont {M.}~\bibnamefont
  {Ohta}}, \bibinfo {author} {\bibfnamefont {S.}~\bibnamefont {Kodama}},
  \bibinfo {author} {\bibfnamefont {Y.}~\bibnamefont {Miyamoto}}, \bibinfo
  {author} {\bibfnamefont {W.}~\bibnamefont {Osten}}, \bibinfo {author}
  {\bibfnamefont {M.}~\bibnamefont {Takeda}},\ and\ \bibinfo {author}
  {\bibfnamefont {E.}~\bibnamefont {Watanabe}},\ }\bibfield  {title} {\bibinfo
  {title} {3d imaging through a highly heterogeneous double-composite random
  medium by common-path phase-shift digital holography},\ }\href@noop {}
  {\bibfield  {journal} {\bibinfo  {journal} {Optics Letters}\ }\textbf
  {\bibinfo {volume} {47}},\ \bibinfo {pages} {1170} (\bibinfo {year}
  {2022})}\BibitemShut {NoStop}%
\bibitem [{\citenamefont {Han}\ \emph {et~al.}(2023)\citenamefont {Han},
  \citenamefont {Peng}, \citenamefont {Li}, \citenamefont {Wang}, \citenamefont
  {Sun},\ and\ \citenamefont {Yao}}]{han2023extending}%
  \BibitemOpen
  \bibfield  {author} {\bibinfo {author} {\bibfnamefont {T.}~\bibnamefont
  {Han}}, \bibinfo {author} {\bibfnamefont {T.}~\bibnamefont {Peng}}, \bibinfo
  {author} {\bibfnamefont {R.}~\bibnamefont {Li}}, \bibinfo {author}
  {\bibfnamefont {K.}~\bibnamefont {Wang}}, \bibinfo {author} {\bibfnamefont
  {D.}~\bibnamefont {Sun}},\ and\ \bibinfo {author} {\bibfnamefont
  {B.}~\bibnamefont {Yao}},\ }\bibfield  {title} {\bibinfo {title} {Extending
  the imaging depth of field through scattering media by wavefront shaping of
  non-diffraction beams},\ }\bibfield  {journal} {\bibinfo  {journal}
  {Photonics}\ }\textbf {\bibinfo {volume} {10}},\ \href
  {https://doi.org/10.3390/photonics10050497} {10.3390/photonics10050497}
  (\bibinfo {year} {2023})\BibitemShut {NoStop}%
\bibitem [{\citenamefont {Harm}\ \emph {et~al.}(2014)\citenamefont {Harm},
  \citenamefont {Roider}, \citenamefont {Jesacher}, \citenamefont {Bernet},\
  and\ \citenamefont {Ritsch-Marte}}]{harm2014lensless}%
  \BibitemOpen
  \bibfield  {author} {\bibinfo {author} {\bibfnamefont {W.}~\bibnamefont
  {Harm}}, \bibinfo {author} {\bibfnamefont {C.}~\bibnamefont {Roider}},
  \bibinfo {author} {\bibfnamefont {A.}~\bibnamefont {Jesacher}}, \bibinfo
  {author} {\bibfnamefont {S.}~\bibnamefont {Bernet}},\ and\ \bibinfo {author}
  {\bibfnamefont {M.}~\bibnamefont {Ritsch-Marte}},\ }\bibfield  {title}
  {\bibinfo {title} {Lensless imaging through thin diffusive media},\
  }\href@noop {} {\bibfield  {journal} {\bibinfo  {journal} {Optics express}\
  }\textbf {\bibinfo {volume} {22}},\ \bibinfo {pages} {22146} (\bibinfo {year}
  {2014})}\BibitemShut {NoStop}%
\bibitem [{\citenamefont {Dalgarno}\ \emph {et~al.}(2012)\citenamefont
  {Dalgarno}, \citenamefont {{\v{C}}i{\v{z}}m{\'a}r}, \citenamefont
  {Vettenburg}, \citenamefont {Nylk}, \citenamefont {Gunn-Moore},\ and\
  \citenamefont {Dholakia}}]{dalgarno2012wavefront}%
  \BibitemOpen
  \bibfield  {author} {\bibinfo {author} {\bibfnamefont {H.~I.~C.}\
  \bibnamefont {Dalgarno}}, \bibinfo {author} {\bibfnamefont {T.}~\bibnamefont
  {{\v{C}}i{\v{z}}m{\'a}r}}, \bibinfo {author} {\bibfnamefont {T.}~\bibnamefont
  {Vettenburg}}, \bibinfo {author} {\bibfnamefont {J.}~\bibnamefont {Nylk}},
  \bibinfo {author} {\bibfnamefont {F.~J.}\ \bibnamefont {Gunn-Moore}},\ and\
  \bibinfo {author} {\bibfnamefont {K.}~\bibnamefont {Dholakia}},\ }\bibfield
  {title} {\bibinfo {title} {Wavefront corrected light sheet microscopy in
  turbid media},\ }\href@noop {} {\bibfield  {journal} {\bibinfo  {journal}
  {Applied Physics Letters}\ }\textbf {\bibinfo {volume} {100}} (\bibinfo
  {year} {2012})}\BibitemShut {NoStop}%
\bibitem [{\citenamefont {Andrews}\ and\ \citenamefont
  {Phillips}(2005)}]{andrews2005laser}%
  \BibitemOpen
  \bibfield  {author} {\bibinfo {author} {\bibfnamefont {L.}~\bibnamefont
  {Andrews}}\ and\ \bibinfo {author} {\bibfnamefont {R.}~\bibnamefont
  {Phillips}},\ }\href {https://books.google.com.mx/books?id=4NXHYg70qqIC}
  {\emph {\bibinfo {title} {Laser Beam Propagation Through Random Media}}},\
  Online access with subscription: SPIE Digital Library\ (\bibinfo  {publisher}
  {Society of Photo Optical},\ \bibinfo {year} {2005})\BibitemShut {NoStop}%
\bibitem [{\citenamefont {Yuan}\ \emph {et~al.}(2021)\citenamefont {Yuan},
  \citenamefont {Xu}, \citenamefont {Zheng}, \citenamefont {Fu}, \citenamefont
  {Wang},\ and\ \citenamefont {Qin}}]{yuan2021experimental}%
  \BibitemOpen
  \bibfield  {author} {\bibinfo {author} {\bibfnamefont {W.}~\bibnamefont
  {Yuan}}, \bibinfo {author} {\bibfnamefont {Y.}~\bibnamefont {Xu}}, \bibinfo
  {author} {\bibfnamefont {K.}~\bibnamefont {Zheng}}, \bibinfo {author}
  {\bibfnamefont {S.}~\bibnamefont {Fu}}, \bibinfo {author} {\bibfnamefont
  {Y.}~\bibnamefont {Wang}},\ and\ \bibinfo {author} {\bibfnamefont
  {Y.}~\bibnamefont {Qin}},\ }\bibfield  {title} {\bibinfo {title}
  {Experimental generation of perfect optical vortices through strongly
  scattering media},\ }\href@noop {} {\bibfield  {journal} {\bibinfo  {journal}
  {Optics Letters}\ }\textbf {\bibinfo {volume} {46}},\ \bibinfo {pages} {4156}
  (\bibinfo {year} {2021})}\BibitemShut {NoStop}%
\bibitem [{\citenamefont {Di~Battista}\ \emph {et~al.}(2016)\citenamefont
  {Di~Battista}, \citenamefont {Ancora}, \citenamefont {Leonetti},\ and\
  \citenamefont {Zacharakis}}]{di2016tailoring}%
  \BibitemOpen
  \bibfield  {author} {\bibinfo {author} {\bibfnamefont {D.}~\bibnamefont
  {Di~Battista}}, \bibinfo {author} {\bibfnamefont {D.}~\bibnamefont {Ancora}},
  \bibinfo {author} {\bibfnamefont {M.}~\bibnamefont {Leonetti}},\ and\
  \bibinfo {author} {\bibfnamefont {G.}~\bibnamefont {Zacharakis}},\ }\bibfield
   {title} {\bibinfo {title} {Tailoring non-diffractive beams from amorphous
  light speckles},\ }\href@noop {} {\bibfield  {journal} {\bibinfo  {journal}
  {Applied Physics Letters}\ }\textbf {\bibinfo {volume} {109}} (\bibinfo
  {year} {2016})}\BibitemShut {NoStop}%
\bibitem [{\citenamefont {Wang}\ \emph {et~al.}(2016)\citenamefont {Wang},
  \citenamefont {Gozali}, \citenamefont {Shi}, \citenamefont {Lindwasser},\
  and\ \citenamefont {Alfano}}]{wang2016deep}%
  \BibitemOpen
  \bibfield  {author} {\bibinfo {author} {\bibfnamefont {W.}~\bibnamefont
  {Wang}}, \bibinfo {author} {\bibfnamefont {R.}~\bibnamefont {Gozali}},
  \bibinfo {author} {\bibfnamefont {L.}~\bibnamefont {Shi}}, \bibinfo {author}
  {\bibfnamefont {L.}~\bibnamefont {Lindwasser}},\ and\ \bibinfo {author}
  {\bibfnamefont {R.}~\bibnamefont {Alfano}},\ }\bibfield  {title} {\bibinfo
  {title} {Deep transmission of laguerre--gaussian vortex beams through turbid
  scattering media},\ }\href@noop {} {\bibfield  {journal} {\bibinfo  {journal}
  {Optics letters}\ }\textbf {\bibinfo {volume} {41}},\ \bibinfo {pages} {2069}
  (\bibinfo {year} {2016})}\BibitemShut {NoStop}%
\bibitem [{\citenamefont {Chen}\ \emph {et~al.}(2018)\citenamefont {Chen},
  \citenamefont {Hu}, \citenamefont {Ji},\ and\ \citenamefont
  {Pu}}]{chen2018needle}%
  \BibitemOpen
  \bibfield  {author} {\bibinfo {author} {\bibfnamefont {Z.}~\bibnamefont
  {Chen}}, \bibinfo {author} {\bibfnamefont {X.}~\bibnamefont {Hu}}, \bibinfo
  {author} {\bibfnamefont {X.}~\bibnamefont {Ji}},\ and\ \bibinfo {author}
  {\bibfnamefont {J.}~\bibnamefont {Pu}},\ }\bibfield  {title} {\bibinfo
  {title} {Needle beam generated by a laser beam passing through a scattering
  medium},\ }\href@noop {} {\bibfield  {journal} {\bibinfo  {journal} {IEEE
  Photonics Journal}\ }\textbf {\bibinfo {volume} {10}},\ \bibinfo {pages} {1}
  (\bibinfo {year} {2018})}\BibitemShut {NoStop}%
\bibitem [{\citenamefont {Popoff}\ \emph {et~al.}(2010)\citenamefont {Popoff},
  \citenamefont {Lerosey}, \citenamefont {Carminati}, \citenamefont {Fink},
  \citenamefont {Boccara},\ and\ \citenamefont {Gigan}}]{T_Mat1}%
  \BibitemOpen
  \bibfield  {author} {\bibinfo {author} {\bibfnamefont {S.~M.}\ \bibnamefont
  {Popoff}}, \bibinfo {author} {\bibfnamefont {G.}~\bibnamefont {Lerosey}},
  \bibinfo {author} {\bibfnamefont {R.}~\bibnamefont {Carminati}}, \bibinfo
  {author} {\bibfnamefont {M.}~\bibnamefont {Fink}}, \bibinfo {author}
  {\bibfnamefont {A.~C.}\ \bibnamefont {Boccara}},\ and\ \bibinfo {author}
  {\bibfnamefont {S.}~\bibnamefont {Gigan}},\ }\bibfield  {title} {\bibinfo
  {title} {Measuring the transmission matrix in optics: An approach to the
  study and control of light propagation in disordered media},\ }\href
  {https://doi.org/10.1103/PhysRevLett.104.100601} {\bibfield  {journal}
  {\bibinfo  {journal} {Phys. Rev. Lett.}\ }\textbf {\bibinfo {volume} {104}},\
  \bibinfo {pages} {100601} (\bibinfo {year} {2010})}\BibitemShut {NoStop}%
\bibitem [{\citenamefont {Liutkus}\ \emph {et~al.}(2014)\citenamefont
  {Liutkus}, \citenamefont {Martina}, \citenamefont {Popoff}, \citenamefont
  {Chardon}, \citenamefont {Katz}, \citenamefont {Lerosey}, \citenamefont
  {Gigan}, \citenamefont {Daudet},\ and\ \citenamefont {Carron}}]{T_Mat3}%
  \BibitemOpen
  \bibfield  {author} {\bibinfo {author} {\bibfnamefont {A.}~\bibnamefont
  {Liutkus}}, \bibinfo {author} {\bibfnamefont {D.}~\bibnamefont {Martina}},
  \bibinfo {author} {\bibfnamefont {S.}~\bibnamefont {Popoff}}, \bibinfo
  {author} {\bibfnamefont {G.}~\bibnamefont {Chardon}}, \bibinfo {author}
  {\bibfnamefont {O.}~\bibnamefont {Katz}}, \bibinfo {author} {\bibfnamefont
  {G.}~\bibnamefont {Lerosey}}, \bibinfo {author} {\bibfnamefont
  {S.}~\bibnamefont {Gigan}}, \bibinfo {author} {\bibfnamefont
  {L.}~\bibnamefont {Daudet}},\ and\ \bibinfo {author} {\bibfnamefont
  {I.}~\bibnamefont {Carron}},\ }\bibfield  {title} {\bibinfo {title} {Imaging
  with nature: Compressive imaging using a multiply scattering medium},\
  }\href@noop {} {\bibfield  {journal} {\bibinfo  {journal} {Scientific
  Reports}\ }\textbf {\bibinfo {volume} {4}},\ \bibinfo {pages} {5552}
  (\bibinfo {year} {2014})}\BibitemShut {NoStop}%
\bibitem [{\citenamefont {Skarsoulis}\ \emph {et~al.}(2021)\citenamefont
  {Skarsoulis}, \citenamefont {Kakkava},\ and\ \citenamefont
  {Psaltis}}]{Skarsoulis:21}%
  \BibitemOpen
  \bibfield  {author} {\bibinfo {author} {\bibfnamefont {K.}~\bibnamefont
  {Skarsoulis}}, \bibinfo {author} {\bibfnamefont {E.}~\bibnamefont
  {Kakkava}},\ and\ \bibinfo {author} {\bibfnamefont {D.}~\bibnamefont
  {Psaltis}},\ }\bibfield  {title} {\bibinfo {title} {Predicting optical
  transmission through complex scattering media from reflection patterns with
  deep neural networks},\ }\href
  {https://doi.org/https://doi.org/10.1016/j.optcom.2021.126968} {\bibfield
  {journal} {\bibinfo  {journal} {Optics Communications}\ }\textbf {\bibinfo
  {volume} {492}},\ \bibinfo {pages} {126968} (\bibinfo {year}
  {2021})}\BibitemShut {NoStop}%
\bibitem [{\citenamefont {Scheidt}\ and\ \citenamefont
  {Quinto-Su}(2023)}]{OurPaper}%
  \BibitemOpen
  \bibfield  {author} {\bibinfo {author} {\bibfnamefont {D.}~\bibnamefont
  {Scheidt}}\ and\ \bibinfo {author} {\bibfnamefont {P.~A.}\ \bibnamefont
  {Quinto-Su}},\ }\bibfield  {title} {\bibinfo {title} {Comparison between
  hadamard and canonical bases for in situ wavefront correction and the effect
  of ordering in compressive sensing},\ }\href
  {https://doi.org/10.1364/JOSAA.473940} {\bibfield  {journal} {\bibinfo
  {journal} {J. Opt. Soc. Am. A}\ }\textbf {\bibinfo {volume} {40}},\ \bibinfo
  {pages} {45} (\bibinfo {year} {2023})}\BibitemShut {NoStop}%
\bibitem [{\citenamefont {McGloin}\ and\ \citenamefont
  {Dholakia}(2005)}]{mcgloin2005bessel}%
  \BibitemOpen
  \bibfield  {author} {\bibinfo {author} {\bibfnamefont {D.}~\bibnamefont
  {McGloin}}\ and\ \bibinfo {author} {\bibfnamefont {K.}~\bibnamefont
  {Dholakia}},\ }\bibfield  {title} {\bibinfo {title} {Bessel beams:
  diffraction in a new light},\ }\href@noop {} {\bibfield  {journal} {\bibinfo
  {journal} {Contemporary physics}\ }\textbf {\bibinfo {volume} {46}},\
  \bibinfo {pages} {15} (\bibinfo {year} {2005})}\BibitemShut {NoStop}%
\bibitem [{\citenamefont {Fahrbach}\ \emph {et~al.}(2013)\citenamefont
  {Fahrbach}, \citenamefont {Gurchenkov}, \citenamefont {Alessandri},
  \citenamefont {Nassoy},\ and\ \citenamefont {Rohrbach}}]{fahrbach2013light}%
  \BibitemOpen
  \bibfield  {author} {\bibinfo {author} {\bibfnamefont {F.~O.}\ \bibnamefont
  {Fahrbach}}, \bibinfo {author} {\bibfnamefont {V.}~\bibnamefont
  {Gurchenkov}}, \bibinfo {author} {\bibfnamefont {K.}~\bibnamefont
  {Alessandri}}, \bibinfo {author} {\bibfnamefont {P.}~\bibnamefont {Nassoy}},\
  and\ \bibinfo {author} {\bibfnamefont {A.}~\bibnamefont {Rohrbach}},\
  }\bibfield  {title} {\bibinfo {title} {Light-sheet microscopy in thick media
  using scanned bessel beams and two-photon fluorescence excitation},\
  }\href@noop {} {\bibfield  {journal} {\bibinfo  {journal} {Optics express}\
  }\textbf {\bibinfo {volume} {21}},\ \bibinfo {pages} {13824} (\bibinfo {year}
  {2013})}\BibitemShut {NoStop}%
\bibitem [{\citenamefont {Fahrbach}\ and\ \citenamefont
  {Rohrbach}(2012)}]{Fahrbach:12}%
  \BibitemOpen
  \bibfield  {author} {\bibinfo {author} {\bibfnamefont {F.}~\bibnamefont
  {Fahrbach}}\ and\ \bibinfo {author} {\bibfnamefont {A.}~\bibnamefont
  {Rohrbach}},\ }\bibfield  {title} {\bibinfo {title} {Propagation stability of
  self-reconstructing bessel beams enables contrast-enhanced imaging in thick
  media},\ }\bibfield  {journal} {\bibinfo  {journal} {Nature Communications}\
  }\textbf {\bibinfo {volume} {3}},\ \href
  {https://doi.org/doi.org/10.1038/ncomms1646} {doi.org/10.1038/ncomms1646}
  (\bibinfo {year} {2012})\BibitemShut {NoStop}%
\bibitem [{\citenamefont {Forbes}(2014)}]{forbes2014laser}%
  \BibitemOpen
  \bibfield  {author} {\bibinfo {author} {\bibfnamefont {A.}~\bibnamefont
  {Forbes}},\ }\href {https://books.google.com.mx/books?id=YErSBQAAQBAJ} {\emph
  {\bibinfo {title} {Laser Beam Propagation: Generation and Propagation of
  Customized Light}}}\ (\bibinfo  {publisher} {CRC Press},\ \bibinfo {year}
  {2014})\BibitemShut {NoStop}%
\bibitem [{\citenamefont {Shen}\ \emph {et~al.}(2022)\citenamefont {Shen},
  \citenamefont {Pidishety}, \citenamefont {Nape},\ and\ \citenamefont
  {Dudley}}]{SelfHealing_review}%
  \BibitemOpen
  \bibfield  {author} {\bibinfo {author} {\bibfnamefont {Y.}~\bibnamefont
  {Shen}}, \bibinfo {author} {\bibfnamefont {S.}~\bibnamefont {Pidishety}},
  \bibinfo {author} {\bibfnamefont {I.}~\bibnamefont {Nape}},\ and\ \bibinfo
  {author} {\bibfnamefont {A.}~\bibnamefont {Dudley}},\ }\bibfield  {title}
  {\bibinfo {title} {Self-healing of structured light: a review},\ }\href
  {https://doi.org/10.1088/2040-8986/ac8888} {\bibfield  {journal} {\bibinfo
  {journal} {Journal of Optics}\ }\textbf {\bibinfo {volume} {24}},\ \bibinfo
  {pages} {103001} (\bibinfo {year} {2022})}\BibitemShut {NoStop}%
\bibitem [{\citenamefont {Mahmoud}\ \emph {et~al.}(2013)\citenamefont
  {Mahmoud}, \citenamefont {Shalaby},\ and\ \citenamefont
  {Khalil}}]{DurninRingBessel}%
  \BibitemOpen
  \bibfield  {author} {\bibinfo {author} {\bibfnamefont {M.~A.}\ \bibnamefont
  {Mahmoud}}, \bibinfo {author} {\bibfnamefont {M.~Y.}\ \bibnamefont
  {Shalaby}},\ and\ \bibinfo {author} {\bibfnamefont {D.}~\bibnamefont
  {Khalil}},\ }\bibfield  {title} {\bibinfo {title} {Propagation of bessel
  beams generated using finite-width durnin ring},\ }\href
  {https://doi.org/10.1364/AO.52.000256} {\bibfield  {journal} {\bibinfo
  {journal} {Appl. Opt.}\ }\textbf {\bibinfo {volume} {52}},\ \bibinfo {pages}
  {256} (\bibinfo {year} {2013})}\BibitemShut {NoStop}%
\bibitem [{\citenamefont {Durnin}\ \emph {et~al.}(1987)\citenamefont {Durnin},
  \citenamefont {Miceli~Jr},\ and\ \citenamefont
  {Eberly}}]{durnin1987diffraction}%
  \BibitemOpen
  \bibfield  {author} {\bibinfo {author} {\bibfnamefont {J.}~\bibnamefont
  {Durnin}}, \bibinfo {author} {\bibfnamefont {J.}~\bibnamefont {Miceli~Jr}},\
  and\ \bibinfo {author} {\bibfnamefont {J.~H.}\ \bibnamefont {Eberly}},\
  }\bibfield  {title} {\bibinfo {title} {Diffraction-free beams},\ }\href@noop
  {} {\bibfield  {journal} {\bibinfo  {journal} {Physical review letters}\
  }\textbf {\bibinfo {volume} {58}},\ \bibinfo {pages} {1499} (\bibinfo {year}
  {1987})}\BibitemShut {NoStop}%
\bibitem [{\citenamefont {Lapointe}(1992)}]{BesselBeams}%
  \BibitemOpen
  \bibfield  {author} {\bibinfo {author} {\bibfnamefont {M.}~\bibnamefont
  {Lapointe}},\ }\bibfield  {title} {\bibinfo {title} {Review of
  non-diffracting bessel beam experiments},\ }\href
  {https://doi.org/https://doi.org/10.1016/0030-3992(92)90082-D} {\bibfield
  {journal} {\bibinfo  {journal} {Optics and Laser Technology}\ }\textbf
  {\bibinfo {volume} {24}},\ \bibinfo {pages} {315} (\bibinfo {year}
  {1992})}\BibitemShut {NoStop}%
\bibitem [{\citenamefont {Vellekoop}\ \emph {et~al.}(2010)\citenamefont
  {Vellekoop}, \citenamefont {Lagendijk},\ and\ \citenamefont
  {Mosk}}]{Vellekoop_2010_localization}%
  \BibitemOpen
  \bibfield  {author} {\bibinfo {author} {\bibfnamefont {I.~M.}\ \bibnamefont
  {Vellekoop}}, \bibinfo {author} {\bibfnamefont {A.}~\bibnamefont
  {Lagendijk}},\ and\ \bibinfo {author} {\bibfnamefont {A.~P.}\ \bibnamefont
  {Mosk}},\ }\bibfield  {title} {\bibinfo {title} {Exploiting disorder for
  perfect focusing},\ }\href {https://doi.org/10.1038/nphoton.2010.3}
  {\bibfield  {journal} {\bibinfo  {journal} {Nature Photonics}\ }\textbf
  {\bibinfo {volume} {4}},\ \bibinfo {pages} {320} (\bibinfo {year}
  {2010})}\BibitemShut {NoStop}%
\bibitem [{\citenamefont {Zupancic}\ \emph {et~al.}(2016)\citenamefont
  {Zupancic}, \citenamefont {Preiss}, \citenamefont {Ma}, \citenamefont
  {Lukin}, \citenamefont {Tai}, \citenamefont {Rispoli}, \citenamefont
  {Islam},\ and\ \citenamefont {Greiner}}]{Zupancic:16}%
  \BibitemOpen
  \bibfield  {author} {\bibinfo {author} {\bibfnamefont {P.}~\bibnamefont
  {Zupancic}}, \bibinfo {author} {\bibfnamefont {P.~M.}\ \bibnamefont
  {Preiss}}, \bibinfo {author} {\bibfnamefont {R.}~\bibnamefont {Ma}}, \bibinfo
  {author} {\bibfnamefont {A.}~\bibnamefont {Lukin}}, \bibinfo {author}
  {\bibfnamefont {M.~E.}\ \bibnamefont {Tai}}, \bibinfo {author} {\bibfnamefont
  {M.}~\bibnamefont {Rispoli}}, \bibinfo {author} {\bibfnamefont
  {R.}~\bibnamefont {Islam}},\ and\ \bibinfo {author} {\bibfnamefont
  {M.}~\bibnamefont {Greiner}},\ }\bibfield  {title} {\bibinfo {title}
  {Ultra-precise holographic beam shaping for microscopic quantum control},\
  }\href {https://doi.org/10.1364/OE.24.013881} {\bibfield  {journal} {\bibinfo
   {journal} {Opt. Express}\ }\textbf {\bibinfo {volume} {24}},\ \bibinfo
  {pages} {13881} (\bibinfo {year} {2016})}\BibitemShut {NoStop}%
\bibitem [{\citenamefont {Goodman}(2020)}]{goodman2020speckle}%
  \BibitemOpen
  \bibfield  {author} {\bibinfo {author} {\bibfnamefont {J.}~\bibnamefont
  {Goodman}},\ }\href {https://books.google.com.mx/books?id=3FvmyAEACAAJ}
  {\emph {\bibinfo {title} {Speckle Phenomena in Optics: Theory and
  Applications}}},\ Press Monographs\ (\bibinfo  {publisher} {SPIE Press},\
  \bibinfo {year} {2020})\BibitemShut {NoStop}%
\bibitem [{\citenamefont {Vellekoop}\ and\ \citenamefont
  {Mosk}(2007)}]{Vellekoop:07}%
  \BibitemOpen
  \bibfield  {author} {\bibinfo {author} {\bibfnamefont {I.~M.}\ \bibnamefont
  {Vellekoop}}\ and\ \bibinfo {author} {\bibfnamefont {A.~P.}\ \bibnamefont
  {Mosk}},\ }\bibfield  {title} {\bibinfo {title} {Focusing coherent light
  through opaque strongly scattering media},\ }\href
  {https://doi.org/10.1364/OL.32.002309} {\bibfield  {journal} {\bibinfo
  {journal} {Opt. Lett.}\ }\textbf {\bibinfo {volume} {32}},\ \bibinfo {pages}
  {2309} (\bibinfo {year} {2007})}\BibitemShut {NoStop}%
\end{thebibliography}%
\end{document}